\documentclass[twocolumn,showpacs,preprintnumbers,amsmath,amssymb]{revtex4}
\usepackage{graphicx}% Include figure files
\usepackage{dcolumn}% Align table columns on decimal point
\usepackage{bm}% bold math

\begin{document}

\title{Nonlinear nonequilibrium quasiparticle relaxation in Josephson junctions}

\author{V. M. Krasnov}
\email{vladimir.krasnov@fysik.su.se}

\affiliation{Department of Physics, Stockholm University, AlbaNova
University Center, SE-10691 Stockholm, Sweden}

\date{\today }

\begin{abstract}
Nonequilibrium electrons in superconductors relax and eventually recombine into Cooper pairs. Relaxation is facilitated by electron-boson interaction and is accompanied by emission of nonequilibrium bosons. Here I solve numerically a full set of nonlinear kinetic balance equations for stacked Josephson junctions, which allows analysis of strongly nonequilibrium phenomena. It is shown that nonlinearity becomes significant already at very small disequilibrium. The following new, essentially nonlinear effects are obtained: (i) At even-gap voltages $V=2n \Delta/e$ $(n=2,3,...)$ nonequilibrium bosonic bands overlap. This leads to enhanced emission of $\Omega = 2 \Delta$ bosons and to appearance of dips in tunnel conductance. (ii) A new type of radiative solution is found at strong disequilibrium. It is characterized by the fast stimulated relaxation of nonequilibrium quasiparticles. A stack in this state behaves as a light emitting diode and directly converts electric power to boson emission, without utilization of the ac-Josephson effect. This leads to very high radiation efficacy and to significant radiative cooling of the stack. The phenomenon can be used for realization of a new type of superconducting cascade laser in the THz frequency range.


\end{abstract}

\maketitle

%\section{Introduction}

Nonequilibrium phenomena in Josephson junctions (JJs) are central for operation of many superconducting devices \cite{Detectors,Cooler,Timofeev} and %, but can be detrimental for Josephson quantum electronics \cite{JosEl}. They 
are important for understanding fundamental properties of superconductors \cite{Ginzburg,Andreev,Barends}. Nonequilibrium quasiparticles (QPs) decay by emitting bosons, with which they are interacting. For example, predominantly phonons are emitted in conventional low-$T_c$ superconductors (LTSC) with electron-phonon coupling mechanism \cite{Phonon,Bron,Dayem,Kinder}. Nonequilibrium boson emission was also observed in high-$T_c$ superconductors (HTSC) \cite{Iguchi,Cascade}, although coupling mechanism in HTSC is not yet confidently known. %and several types of bosons may contribute cooperatively to electron pairing \cite{Dolgov,Suppl}.
Analysis of the content of such emission provides a possibility for direct determination of the coupling mechanism in HTSC \cite{Cascade}.

Nonequilibrium effects in stacked intrinsic Josephson junctions (IJJs), naturally formed in layered HTSC such as Bi$_2$Sr$_2$CaCu$_2$O$_{8+\delta}$ (Bi-2212), can be much larger than in LTSC JJs \cite{Suppl}. This is caused (i) by the atomic thickness of CuO$_2$ planes with a very small total density of states (DoS); and by stacking of junctions which (ii) prevents QP leakage from electrodes and (iii) leads to cascade amplification of boson population upon sequential QP tunneling \cite{Cascade}. The two latter factors are well known in semiconducting heterostructures and lead to a $10^4$-fold decrease of the threshold current density in superlattice lasers, compared to single $p$-$n$ junction lasers \cite{Alferov}. I argue that a similar dramatic enhancement of nonequilibrium effects could occur in stacked IJJs and may facilitate population inversion required for lasing \cite{Cascade,Suppl}. The IJJ laser could provide a tunable emission in the frequency range $\sim 0.3$-7 THz and is considered as one of the candidates for filling the ``THz gap" between microwave and infrared technologies \cite{Ozyuzer,Wang}.
%Compact THz sources are demanded for a great variety of applications such as medical imaging, security, industrial and environmental control, telecommunication, e.t.c, as well as for fundamental research e.g. in astronomy, condensed matter physics and biology.A specific significance of the THz frequency range is due to the majority of important molecular transitions (e.g. water in living cells or carbon dioxide in atmosphere) being concentrated in this frequency range.

Understanding LTSC device performance at low $T$ and boson emission from IJJs requires analysis of strongly nonequilibrium states. The appropriate microscopic formalism was developed in the BCS work \cite{BCS}. It represents a set of two coupled nonlinear kinetic equations for QPs and bosons, together with a self-consistency equation for the energy gap $\Delta$. Although, those equations were actively studied before \cite{Detectors,Dayem,BRT,Kaplan,Chang}, previous studies were limited to a simplified version, in which nonequilibrium distribution functions of QPs, $f=F+\delta f$, and bosons, $g=G+\delta g$, were linearized with respect to equilibrium values $F$ and $G$.
%$F(E,T)=[\exp(E/k_BT)+1]^{-1}$ and $G(\Omega,T)=[\exp(\Omega/k_BT)-1]^{-1}$.
As will be shown below, such linearization is valid only for extremely small disequilibrium, $\delta f\ll F$, $\delta g \ll G$, which is hardly satisfied at low enough $T$ or high enough energy, where $F, G\rightarrow 0$. 

In this work I present numerical solutions of the full set of nonlinear equations, with a focus on strongly nonequilibrium phenomena in stacked JJs at high bias. A new type of radiative solution is found, resembling lasing in semiconducting heterostructures. %The following new, essentially nonlinear effects are obtained: (i) Peculiarities in the tunnel conductance appear at even-gap voltages $V=2n \Delta/e$ $(n=2,3,...)$. They are caused by collision of two nonequilibrium bosonic bands and lead to enhanced boson emission at $\Omega \geqslant 2\Delta$. (ii) A novel radiative solution is found in the strongly nonequilibrium case. It is characterized by fast stimulated relaxation of nonequilibrium QPs, accompanied by enhanced emission of low frequency bosons. A junction in this state behaves like an injection laser diode with high luminous efficacy \cite{LED}. % and effectively converts electric power into boson emission, instead of heat.

I consider two voltage biased stacked JJs. The role of the second JJ is to enhance nonequilibrium QP population by preventing QP leakage from the middle electrode, just like in semiconducting double heterostructures \cite{Alferov}. On the other hand, bosons are not confined and can freely escape through JJs. %with the rate proportional to the impingement rate on the barrier.
The boson escape rate and the QP injection rate, relative to the QP relaxation rate in electrodes, are described by coefficients $\gamma_U$ and $\gamma_I$, respectively. All characteristics are shown for the middle electrode. Exact formalism, details of the numerical procedure and estimation of parameters for Bi-2212 IJJs can be found in the Supplementary \cite{Suppl}.

\begin{figure}
\includegraphics[width=2.7in]{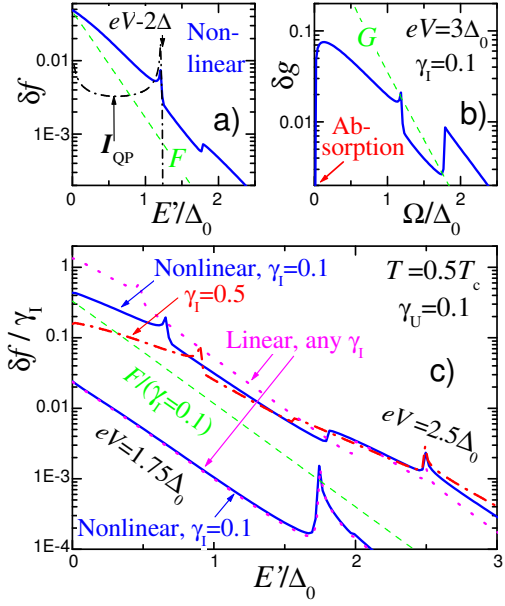}
\caption{\label{Fig_dfdgNonLin} (Color online). QP (a) and boson (b) spectra for the ordinary absorptive solution at $eV=3.0\Delta_0$. Solid lines represent nonequilibrium $\delta f$, $\delta g$, dashed lines-equilibrium $F$, $G$. The dashed-dotted line in (a) shows the QP injection rate. It is seen that there is net accumulation of QPs and absorption of bosons at low energies. c) Comparison of QP spectra, normalized by the injection rate $\gamma_I$, for linear (dotted lines) and nonlinear solutions for $V=1.75\Delta_0/e <V_g$ and $V=2.5\Delta_0/e >V_g$ and for $\gamma_I=0.1$ and $\gamma_I=0.5$ (the red dashed-dotted line, for $eV=2.5\Delta_0$ only). At $eV=1.75\Delta_0$, $\delta f \ll F$ and linear and nonlinear solutions coincide. However at large bias $eV=2.5\Delta_0$, $\delta f > F$ and linear solution becomes inaccurate. Nonlinearity stimulates QP decay, as seen from the strong decrease of $\delta f(0)/\gamma_I$ with $\gamma_I$.  }
\end{figure}

The QP current in JJs rises stepwise at the sum-gap voltage $V_g = 2 \Delta /e$. %(see the inset in Fig. \ref{Fig_IVCsol12}). 
At $V>V_g$ it has two sharp maxima at $E'=0$ and $E'=eV-2\Delta$, as shown by the dashed-dotted line in Fig. \ref{Fig_dfdgNonLin} a). Here $E'=E-\Delta$ is counted from the edge of the gap. Injected QPs relax via spontaneous and stimulated emission of bremsstrahlung bosons and eventually recombine into Cooper pairs with emission of recombination bosons \cite{Bron,Dayem,Kinder,Cascade}. Opposite processes of boson absorption and pair-breaking slow down the QP decay. The rates of all those processes are proportional to electron-boson spectral function \cite{BCS}, i.e., QP decay is facilitated by electron-boson interaction.

Fig. \ref{Fig_dfdgNonLin} represents full nonlinear solutions for a) QP and b) boson spectra at $V=3\Delta_0/e$, $T = 0.5 T_c$ and for QP injection and boson escape coefficients $\gamma_I = \gamma_U = 0.1$. In Fig. \ref{Fig_dfdgNonLin} c) linear and nonlinear solutions, normalized by $\gamma_I$, are compared. At low bias $V=1.75\Delta_0/e < V_g$ the QP current is small and $\delta f \ll F$. In this case linearization is valid and both solutions coincide. However, at $V = 2.5 \Delta_0 /e > V_g$ the QP current increases dramatically so that $\delta f > F$ and linearization becomes invalid. The red dashed-dotted line in c) represents the full solution for a larger QP injection rate $\gamma_I=0.5$ at $eV=2.5\Delta_0$. It is seen that $\delta f/\gamma_I$ at the same $eV=2.5\Delta_0$ decreases substantially with increasing $\gamma_I$ and disequilibrium, in contrast to the linear solution for which $\delta f/\gamma_I$ is independent of $\gamma_I$. For $\gamma_I=0.5$, $\delta f(0)/\gamma_I $ is an order of magnitude {\it smaller} than for the linear solution, implying that nonlinearity tends to stimulate QP decay.

\begin{figure}
\includegraphics[width=2.7in]{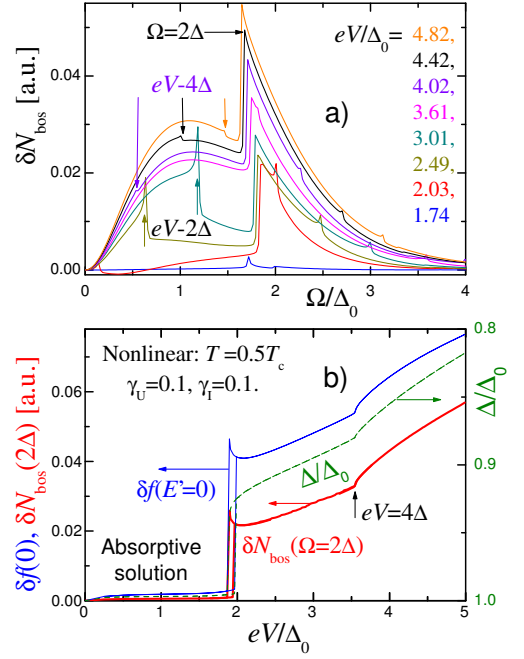}
\caption{\label{Fig_Nph2Dg01} (Color online). (a) Nonlinear boson emission spectra at different $V$. Bremsstrahlung $0 \leqslant \Omega \leqslant eV-2\Delta$ and recombination $\Omega \geqslant 2\Delta$ bands with sharp maxima at band edges are clearly seen for $eV = 2.49$ and $3.01 \Delta_0$. (b) Bias dependence of $\delta f(E'=0)$ (thin line), $\Delta$ (dashed line, right axis, note the reversed scale) and the number of recombination bosons $\delta N_{bos} (\Omega = 2 \Delta)$ (thick line). %Inset in (a) shows $\delta g (\Omega = 0)$ its negative value indicates a net absorption of low-energy bosons.
Step-like increase of $\delta N_{bos}$ occurs at $eV=4\Delta \simeq 3.54 \Delta_0$ upon collision of bremsstrahlung and recombination bosonic bands. This also leads to appearance of secondary QPs and the secondary bremsstrahlung band with $0<\Omega<eV-4\Delta$ (downward arrows in a).}
\end{figure}

Fig. \ref{Fig_Nph2Dg01} a) shows boson emission spectra at different bias. %and for $\gamma_I=0.1$ and the rest of the parameters the same as in Fig. \ref{Fig_dfdgNonLin}.
Bremsstrahlung and recombination bands are clearly seen at $eV=2.49$ and $3.01\Delta_0$. Sharp maxima at the band edges $\Omega = eV-2\Delta$ (upward arrows) and $\Omega = 2\Delta$ indicate the most probable QP decay scenario \cite{Bron}: first QPs fall to the edge of the gap and then recombine therefrom into Cooper pairs, emitting $\Omega = eV-2\Delta$ and  $\Omega = 2\Delta$ bosons, respectively. %Predominance of those processes is due to BCS singularity in the QP DoS.

%\subsection{Nonequilibrium suppression of the gap}
Nonequilibrium QPs suppress $\Delta$ from its equilibrium value $\Delta_0(T)$. In Fig. \ref{Fig_Nph2Dg01} a) this is seen as a gradual shift of the recombination band edge $\Omega=2\Delta$ with $V$. According to the self-consistency equation, the gap is mostly affected by QPs at $E'=0$. Indeed, from Fig. \ref{Fig_Nph2Dg01} b) it is seen that $V$-dependence of $\Delta$ follows $\delta f(0)$. Note a hysteresis upon sweeping through $V_g$. It is caused by nonlinearity of the self-consistency equation.

\begin{figure}
\includegraphics[width=2.7in]{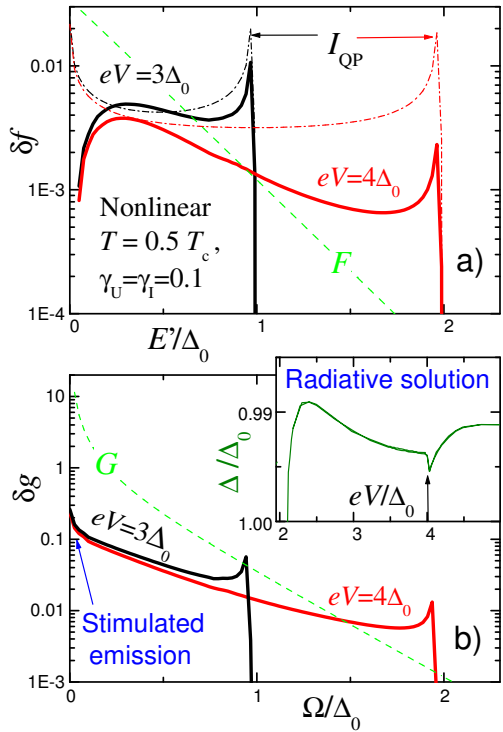}
\caption{\label{Fig_Sol2_all} (Color online). QP (a) and boson (b) spectra for the radiative solution. Presentation is similar to Figs. 1 a,b). Reduction of $\delta f(0)$ at the edge of the gap in (a) indicates accelerated relaxation due to stimulated boson emission. The corresponding increase of low-frequency boson population $\delta g(0)$ is indicated  in (b). Inset in (b) shows $V$-dependence of $\Delta$: gap suppression is very small due to small $\delta f(0)$.}
\end{figure}

Despite substantial quantitative differences, the nonlinear solution described above is
qualitatively similar to the linear solution. From comparison of QP injection rates $I_{QP}(E)$ and $\delta f (E)$ in Fig. \ref{Fig_dfdgNonLin} a) it is seen that the amount QPs is reduced with respect to $I_{QP}$ at high energies $E' \simeq eV-2\Delta$ as a result of QP relaxation, but is largely enhanced at $E'\simeq 0$. Accumulation of QPs at the edge of the gap occurs not only because recombination is the only sink channel for QPs in the considered case, but also because QP relaxation is {\it slowed down} by reabsorption of bosons. As a result there is a net absorption of bosons at low $\Omega$, $\delta g(0)<0$. Therefore, I will refer to this type of solution as the ``absorptive" solution.

\begin{figure}
\includegraphics[width=2.7in]{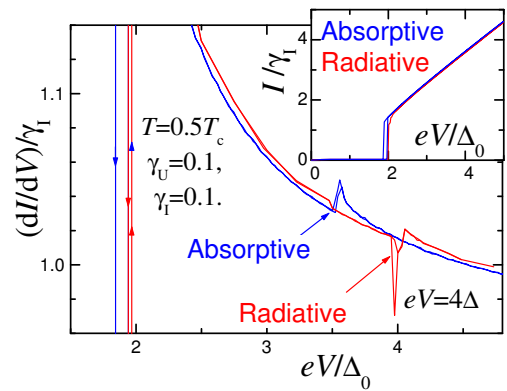}
\caption{\label{Fig_IVCsol12} (Color online). Normalized tunnel conductance vs. bias voltage for absorptive (blue) and radiative (red) solutions and for the same junction parameters. The double-gap dip/step occurs in both cases at $V=2V_g=4\Delta/e$. However, the gap is much stronger suppressed for the absorptive solution, despite similar dissipation powers along the $I-V$ curves, shown in the inset. Small gap suppression in the radiative state is due to efficient radiative cooling. }
\end{figure}

With increasing disequilibrium, above some threshold QP injection rate, the absorptive solution becomes unstable and switches to a different one, shown in Fig. \ref{Fig_Sol2_all} for the same parameters as in Figs. \ref{Fig_dfdgNonLin} and \ref{Fig_Nph2Dg01}. In this case there is no significant accumulation of QPs at $E'=0$. As a consequence, suppression of $\Delta$ is very small, as shown in the inset. Instead there is a large net emission of low-energy bosons, marked by an upward arrow in Fig. \ref{Fig_Sol2_all} b). Apparently, relaxation of QPs is {\it accelerated} by stimulated emission of bremsstrahlung bosons (note that even for the absorptive solution the nonlinearity accelerated QP decay, see Fig. \ref{Fig_dfdgNonLin} c). Significance of stimulated emission is obvious from the large total boson population $g(\Omega \rightarrow 0) \gg 1$. Therefore, I will refer to this new type of solution as the ``radiative" solution. Once achieved, the radiative solution is sustained even at lower bias, typically down to $V_g$, as is the case in Fig. \ref{Fig_Sol2_all}. I want to emphasize that existence of two solutions is the consequences of nonlinearity of kinetic equations.

Upward arrows in Fig. \ref{Fig_Nph2Dg01} b) and the inset in Fig. \ref{Fig_Sol2_all} b) indicate a singularity at the double-gap voltage $V = 2V_g = 4\Delta/e$. At this bias bremsstrahlung and recombination bands overlap, as seen from spectra at $eV>3.54\Delta_0$ in Fig. \ref{Fig_Nph2Dg01} a). This leads to:

A) {\it Appearance of secondary QPs and a secondary bremsstrahlung band}: at $V>2V_g$ high energy bremsstrahlung bosons become capable of breaking Cooper pairs. This leads to appearance of secondary QPs with energies $0\leqslant E'\leqslant eV-4\Delta$ \cite{Bron}. The corresponding step-like increase of $\delta f(0)$ is clearly seen in Fig. \ref{Fig_Nph2Dg01} b). Relaxation of secondary QPs leads to appearance of secondary bremsstrahlung bosons with $0\leqslant \Omega \leqslant eV-4\Delta$, marked by downward arrows in Fig. \ref{Fig_Nph2Dg01} a). Recombination of secondary QPs leads to enhanced emission of $\Omega = 2 \Delta$ bosons. At $eV>4\Delta$ the amount of such bosons increase stepwise, as clearly seen in Fig. \ref{Fig_Nph2Dg01} b).

B) {\it Double-gap singularity in $dI/dV$}: Step-like increase of $\delta f(0)$ leads to the corresponding step-like decrease of $\Delta(V)$, see Fig. \ref{Fig_Nph2Dg01} b). This in turn leads to appearance of a dip/step in tunneling conductance
as demonstrated in Fig. \ref{Fig_IVCsol12}. Note that the feature is sharper for the radiative solution due to sharper step in $\Delta(V)$, despite smaller absolute gap suppression. Similar features occur at all even-gap voltages $eV=2n\Delta$, $(n=2,3...)$ due to collision between recombination and secondary bremsstrahlung bands \cite{Bron}. 

Noticeably, $I-V$ curves for both solutions are very similar, as shown by the inset in Fig. \ref{Fig_IVCsol12}. The large difference in $\Delta (V)$ for absorptive and radiative solutions for similar dissipative powers $P=IV$ emphasizes inapplicability of the concept of thermal conductivity $\kappa$ for description of strongly nonequilibrium phenomena in IJJs \cite{Heating}. In superconductors $\kappa$ rapidly freezes out at $T\rightarrow 0$. Numerical simulations presented here correspond to $\kappa = 0$ because there is no diffusive escape neither for QPs nor bosons. Nevertheless, neither QP nor boson sub-systems are strongly overheated. This occurs because electric power is effectively (with $100\%$ conversion efficiency) radiated from JJs via ballistic boson emission. The phenomenon is well known as radiative cooling in light emitting diodes \cite{Alferov}.

I want to emphasize similarities between nonlinear nonequilibrium phenomena in stacked JJs and semiconducting heterostructures \cite{Alferov}, especially Quantum Cascade Lasers \cite{Cascade,QCL}. The absorptive and radiative solutions, obtained here, are similar to light emitting and lasing states in injection diodes. Lasing in modern diodes occurs above a threshold $J_{th} \sim 10-100$ A/cm$^2$ at room temperature and $J_{th}$ decreases exponentially with decreasing $T$ \cite{Alferov}. For IJJs, $J(V=V_g) \simeq 10^3$-$10^4$ A/cm$^2$ at much lower $T = 4.2K$ \cite{Doping,Heating}. According to preliminary results (not shown) the threshold for switching from the absorptive to the radiative state in JJs also strongly decreases with $T$. This is caused by the exponential dependence $F(T)$ which makes it easier to achieve strong nonlinearity $\delta f \gg F$ at low $T$. Furthermore, large Bi-2212 mesa structures used in emission experiments \cite{Ozyuzer,Wang} may contain up to 1000 stacked IJJs, which should result in significant cascade amplification of boson population \cite{Cascade}. Therefore, extreme disequilibrium (population inversion) can be achieved in such mesas.

Obtained results are consistent with several experimental observations: 
(i) I have shown that QP relaxation becomes nonlinear when $\delta f \gtrsim F$ (see Fig. 1 c). Since $F$ freezes out exponentially with decreasing $T$, QP relaxation inevitably becomes nonlinear at low enough $T$. In this case recombination occurs between two nonequilibrium QPs rather than one nonequilibrium and one equilibrium QP, recombination time becomes independent of $T$ and dependent of QP injection rate, as observed in LTSC JJs at mK temperatures \cite{Timofeev,Barends}.
(ii) Similar dips/steps in $dI/dV$ at even-gap voltages (see Fig. 4) were observed for small Bi-2212 IJJs \cite{Cascade}. Abrupt appearance of secondary QPs at those voltages (see Fig. 2 b) was directly observed in LTSC JJs \cite{Phonon,Dayem,Bron}.
(iii) Appearance of radiative cooling upon establishing the radiative state in IJJs is consistent with the observed small suppression of $\Delta$ \cite{Cascade} and strong reduction of self-heating at high bias \cite{Heating}. The radiative state can be also responsible for recently observed laser-like photon emission from large Bi-2212 mesa structures \cite{Ozyuzer}. In that case the mesa itself acts as a Fabry-Perot resonator, selecting cavity (Fiske) modes \cite{Wang,Fiske}.

In conclusion, solution of the full nonlinear set of kinetic balance equations allowed analysis of arbitrary strong nonequilibrium phenomena in Josephson junctions. This is important for quantitative analysis of superconducting devices at low $T$ and for understanding of nonequilibrium boson emission from IJJs. The discovered radiative state indicates a possibility of realization of a new type of superconducting cascade laser (SCL) \cite{Cascade}. Unlike the existing Josephson oscillators which utilize the ac-Josephson effect for conversion of electric power into radiation \cite{Koshelets}, the SCL is based on direct conversion of electric power into boson emission via non-equilibrium QP relaxation upon sequential tunneling in stacked junctions. The mechanism is similar to lasing in semiconducting heterostructures \cite{Alferov,QCL} and allows very high radiation efficiency. The advantage of employing superconductors instead of semiconductors is in the smaller energy gap, which facilitates operation in the important THz frequency range.

%\begin{acknowledgments}
Financial support from the K.{\&}A. Wallenberg foundation, the Swedish Research Council and the SU-Core Facility in Nanotechnology is gratefully acknowledged.
%\end{acknowledgments}

%\end{multicols}

\begin{references}

\bibitem{Detectors} A.G. Kozorezov {\em et al.}, {\em Phys. Rev. B} {\bf 77}, 014501 (2008);
A.A. Golubov {\em et al.}, {\em ibid.}, %Phys. Rev. B}
{\bf 49}, 12953 (1994).

\bibitem{Cooler} A.M. Clark {\em et al.}, {\em Appl. Phys. Lett.} {\bf 86}, 173508 (2005).
    
\bibitem{Timofeev} A.V. Timofeev {\em et al.}, {\em Phys. Rev. Lett.} {\bf 102}, 200801 (2009).

%\bibitem{JosEl} S. Intiso {\em et al.}, {\em Supercond. Sci. Technol.} {\bf 19}, S335 (2006). %A.M. Savin et al., {\em Appl. Phys. Lett.} {\bf 89}, 133505 (2006)
%\bibitem{Nernst} Nernst

\bibitem{Ginzburg} V.L. Ginzburg, {\em Usp. Fiz. Nauk} {\bf 174}, 1240 (2004).

\bibitem{Andreev} M. Octavio {\em et al.}, %, M.Tinkham, G.E.Blonder and T.M. Klapwijk,
{\em Phys. Rev. B} {\bf 27}, 6739 (1983)

\bibitem{Barends} R. Barends {\em et al.}, {\em Phys. Rev. Lett.} {\bf 100}, 257002 (2008).

\bibitem{Phonon} W. Eisenmenger and A.H. Dayem, {\em Phys. Rev. Lett.} {\bf 18}, 125
(1967); R.C. Dynes and V. Narayanamurti, {\em Phys. Rev. B} {\bf 6}, 143 (1972). %; P. Berberich, R. Buemann, and H. Kinder, {\em Phys. Rev. Lett.} {\bf 49}, 1500 (1982).

\bibitem{Dayem} A.H. Dayem and J.J. Wiegand, {\em Phys.Rev.B} {\bf 5}, 4390 (1972)

\bibitem{Bron} W.E. Bron, {\em Rep. Prog. Phys.} {\bf 43}, 20 (1980)

%\bibitem{Dynes} R.C. Dynes and V. Narayanamurti, {\em Phys. Rev. B} {\bf 6}, 143 (1972)

\bibitem{Kinder} H. Kinder, {\em Phys. Rev. Lett.} {\bf 28}, 1564 (1972).

\bibitem{Iguchi} I. Iguchi {\em et al.}, {\em Phys. Rev. B} {\bf 61}, 689 (2000).

\bibitem{Cascade} V.M. Krasnov, {\em Phys. Rev. Lett.} {\bf 97}, 257003 (2006).

%\bibitem{Dolgov} M.L. Kulic and O.V. Dolgov, {\em Phys.Rev.B} {\bf 76}, 132511 (2007).

\bibitem{Suppl} See the EPAPS Document No. ??? for details of numerical simulations. For more information on EPAPS, see http://www.aip.org/pubservs/epaps.html.

\bibitem{Alferov} Z.I. Alferov, {\em Rev. Mod. Phys.} {\bf 73}, 767 (2001).
%\bibitem{LED} O.Heikkil\"{a}, J.Oksanen, and J.Tulkki, J.Appl.Phys. 105, 093119 (2009).
%S.Nizamoglu and H.V.Demir, J. Appl. Phys. 105, 083112 (2009)


\bibitem{Ozyuzer} L. Ozyuzer {\em et al.}, {\em Science} {\bf 318}, 1291 (2007).

\bibitem{Wang} H.B. Wang {\em et al.}, {\em Phys. Rev. Lett.} {\bf 102}, 017006 (2009).


\bibitem{BCS} J. Bardeen, L. Cooper and J. Schrieffer, {\em Phys. Rev.} {\bf 108}, 1175 (1957)

\bibitem{BRT} J. Bardeen, G. Rickaysen and L. Tewordt, {\em Phys. Rev.} {\bf 113}, 982 (1959).

\bibitem{Kaplan} S.B. Kaplan {\em et al.}, {\em Phys. Rev. B} {\bf 14}, 4854 (1976).

\bibitem{Chang} J.J. Chang and D.J. Scalapino, {\em Phys.Rev.B.} {\bf 15}, 2651 (1977).
%; ibid 21, 2045 (1980).

\bibitem{Heating} V.M. Krasnov, {\em Phys. Rev. B} {\bf 79}, 214510 (2009);
V.M. Krasnov, M. Sandberg and I. Zogaj, {\em Phys. Rev. Lett.} {\bf 94}, 077003 (2005).

\bibitem{QCL} V. Spagnolo {\em et al.} {\em Appl. Phys. Lett.} {\bf 80}, 4303 (2002).
%J. Faist, et al., {\em Science} {\bf 264}, 553 (1994).

\bibitem{Doping} V.M. Krasnov, {\em Phys. Rev. B} {\bf 65}, 140504(R) (2002).

\bibitem{Fiske} V.M. Krasnov {\em et al.}, %N.Mros, A.Yurgens, and D.Winkler,
{\em Phys. Rev. B} {\bf 59}, 8463 (1999).

\bibitem{Koshelets} V.P. Koshelets and S.V. Shitov, {\em Supercond.Sci.Technol.} {\bf 13}, R53 (2000).

\end{references}
\end{document}